\documentclass[prx, preprintnumbers, amsmath, amssymb, superscriptaddress, twocolumn]{revtex4-1}
\usepackage{graphicx, color, dcolumn, natbib, bm, amssymb, amsmath,color}
\graphicspath{{fig/}}
\usepackage[final=true]{hyperref}
\usepackage[normalem]{ulem}

\begin{document}
\title{Giant enhancement of the skyrmion stability in \\ a chemically strained helimagnet}
\author{A. S. Sukhanov}\thanks{These authors contributed equally.}
\affiliation{Max Planck Institute for Chemical Physics of Solids, D-01187 Dresden, Germany}
\affiliation{Institut f{\"u}r Festk{\"o}rper- und Materialphysik, Technische Universit{\"a}t Dresden, D-01069 Dresden, Germany}
\author{Praveen Vir}\thanks{These authors contributed equally.}
\affiliation{Max Planck Institute for Chemical Physics of Solids, D-01187 Dresden, Germany}
\author{A. Heinemann}
\affiliation{German Engineering Materials Science Centre (GEMS) at Heinz Maier-Leibnitz Zentrum (MLZ), Helmholtz-Zentrum Geesthacht GmbH, D-85747 Garching, Germany}
\author{S. E. Nikitin}
\affiliation{Max Planck Institute for Chemical Physics of Solids, D-01187 Dresden, Germany}
\author{D. Kriegner}
\affiliation{Max Planck Institute for Chemical Physics of Solids, D-01187 Dresden, Germany}
\author{H. Borrmann}
\affiliation{Max Planck Institute for Chemical Physics of Solids, D-01187 Dresden, Germany}
\author{C. Shekhar}
\affiliation{Max Planck Institute for Chemical Physics of Solids, D-01187 Dresden, Germany}
\author{C. Felser}
\affiliation{Max Planck Institute for Chemical Physics of Solids, D-01187 Dresden, Germany}
\author{D. S. Inosov}
\affiliation{Institut f{\"u}r Festk{\"o}rper- und Materialphysik, Technische Universit{\"a}t Dresden, D-01069 Dresden, Germany}
\date{\today}
\begin{abstract}

We employed small-angle neutron scattering to demonstrate that the magnetic skyrmion lattice can be realized in bulk chiral magnets as a thermodynamically stable state at temperatures much lower than the ordering temperature of the material. This is in the regime where temperature fluctuations become completely irrelevant to the formation of the topologically non-trivial magnetic texture. In this attempt we focused on the model helimagnet MnSi, in which the skyrmion lattice was previously well characterized and shown to exist only in a very narrow phase pocket close to the Curie temperature of 29.5~K. We revealed that large uniaxial distortions caused by the crystal-lattice strain in MnSi result in stabilization of the skyrmion lattice in magnetic fields applied perpendicular to the uniaxial strain at temperatures as low as 5~K. To study the bulk chiral magnet subjected to a large uniaxial stress, we have utilized $\mu$m-sized single-crystalline inclusions of MnSi naturally found inside single crystals of the nonmagnetic material Mn$_{11}$Si$_{19}$. The reciprocal-space imaging allowed us to unambiguously identify the stabilization of the skyrmion state over the competing conical spin spiral.

\end{abstract}

\maketitle

Early experimental observations of the skyrmion lattice---a topologically protected spin texture---demonstrated its very limited stability due to a fragile balance of the relevant magnetic interactions in a real material. First found in a bulk sample of the chiral helimagnet MnSi, the skyrmion-lattice (SkL) phase was shown to exist in a narrow region of the temperature\,--\,magnetic-field phase diagram: approximately 2~K wide (much smaller than the Curie temperature $T_{\text C}$ = 29.5 K) in temperature and within the range of 0.1~T in applied field~\cite{Muehlbauer}. Later discoveries of the bulk SkL in other compounds with the chiral space group $P2_13$ \cite{Muehlbauer,Muenzer,Grigoriev07,Pfleiderer,Bauer,Grigoriev09,Seki12,Adams} or $P$4$_1$32 \cite{Tokunaga} supported the same characteristic---the skyrmions existed only in a small phase pocket close to $T_{\text C}$.

Interestingly, the first real-space observation of the skyrmion texture on thin plates of Fe$_{0.5}$Co$_{0.5}$Si revealed significantly extended stability of the SkL phase~\cite{Yu}. The observations were supported by simulations that assumed the 2D character of the system---the approximation relevant when the sample thickness is smaller than the skyrmion size. Butenko \textit{et al}.~\cite{Butenko} theoretically considered the influence of uniaxial magnetic anisotropy on the stability of the SkL in cubic noncentrosymmetric ferromagnets. It was shown, that the induced anisotropy, if strong enough, reduces the energy of the SkL over the competing conical state in finite magnetic fields~\cite{Butenko}. The same theoretical approach was applied to interpret the results of magnetic measurements performed on thin epitaxial films of MnSi~\cite{Karhu,Wilson12,Wilson14}, where some indication of an extended skyrmion stability was noticed. In another study~\cite{Gallagher}, thin films of the $B20$-type compound FeGe, grown on Si(111) substrates, demonstrated a pronounced topological Hall effect at low temperatures far below the SkL phase of the bulk FeGe. Such a magnetotransport phenomenon is expected in the presence of a topologically non-trivial spin structure and may indicate skyrmions with enhanced stability. However, no skyrmion phase was found in FeGe/MgO films~\cite{Zhang}.

Despite a number of reports that the SkL phase possesses extended stability in thin epitaxial films of chiral magnets, only little microscopic evidence was presented up to date~\cite{Meynell}. The discussion of the skyrmion stability in relation to thin films is complicated for the following reasons. On the one hand, all the thin films studied so far have significant strain caused by the lattice mismatch between the substrate and the film~\cite{Karhu,Sinha,Figueroa}. On the other hand, the typical thickness of the films is of the order of the helical modulation length, therefore the contribution from surface effects might be significant. Moreover, both strain and the surface contributions vary with the film thickness, which makes it difficult to disentangle the influence of these mechanisms. Results of Ref.~\cite{Yu_2} obtained on a free-standing thin plate of FeGe pointed to the importance of the surface, as the SkL phase extended to lower temperatures upon decreasing the sample thickness. The surface-induced formation of skyrmions was also identified via a mechanism of chiral surface twist~\cite{Leonov}. Whilst microfabricated strain-free thin plates of skyrmion-hosting materials can be used to study surface-related phenomena, bulk strained crystals must be employed to elucidate the influence of the distortion.

\begin{figure}[t]
        \begin{minipage}{0.99\linewidth}
        \center{\includegraphics[width=1\linewidth]{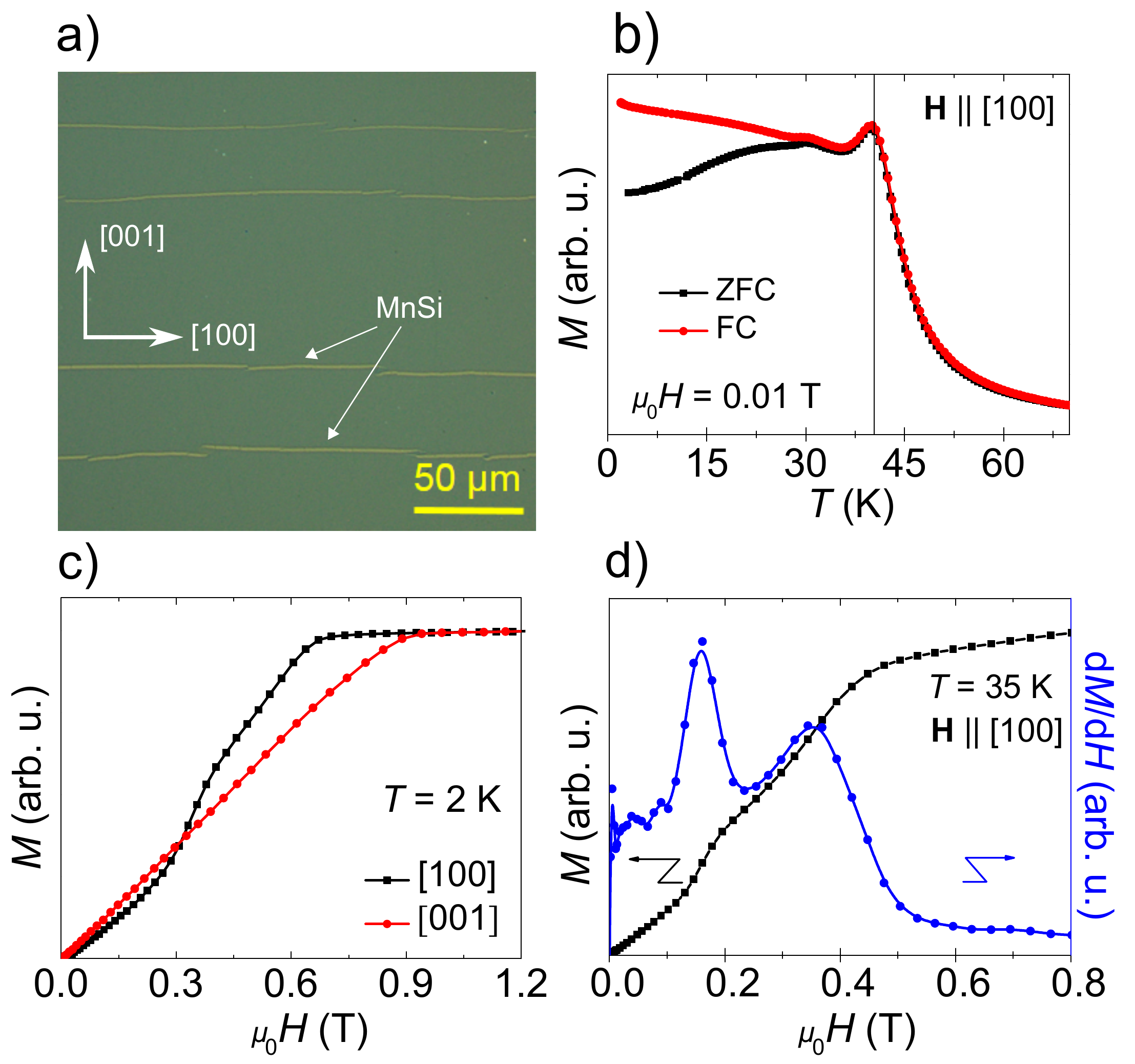}}
        \end{minipage}
        \caption{(color online). (a) The micrograph of MnSi lamellae embedded into the nonmagnetic Mn$_{11}$Si$_{19}$ matrix. (b) ZFC and FC magnetization as a function of temperature; (c) magnetization as a function of the field at $T = 5$~K; (d) metamagnetic transitions at 35 K, solid lines are guides for the eyes.}
        \label{ris:fig1}
\end{figure}

There has been recent progress in understanding the consequences of strain in bulk $B20$ crystals~\cite{Chacon,Nii,Fobes}. Comprehensive small-angle neutron scattering (SANS) measurements~\cite{Chacon} of MnSi under the applied uniaxial pressure showed the twofold enhancement (reduction) of the skyrmion stability when the field was applied perpendicular (parallel) to the strain direction. Microwave absorption and resonant x-ray experiments on the insulating chiral magnet Cu$_2$OSeO$_3$ subjected to a tensile strain also showed the same trend~\cite{Seki17,Okamura}.

Nevertheless, it remained unclear if the skyrmions can be stabilized in chiral magnets in the whole temperature range below $T_{\text C}$. The uniaxial distortions caused by the controlled hydrostatic pressure or tensile strain are an order of magnitude smaller than the characteristic strains in epitaxial thin films~\cite{Karhu,Chacon,Seki17}. That also complicates the identification, which mechanism (surface or strain) plays the leading role in potential enhancement of the skyrmion stability in thin films.

In the present paper we report the results of studying the model chiral helimagnet, MnSi, which is chemically strained to the values typically found in the thin films while kept in the bulk form. Using SANS we were able to unambiguously interpret all the magnetic phases (helical, conical, SkL, field-polarized) for different orientation of the applied field ($H \perp \sigma$ and $H \parallel \sigma$) and span the whole $T$-$H$ parameter space to construct the full phase diagram. The reciprocal-space imaging allowed us to identify the SkL explicitly and investigate its temperature stability under high uniaxial distortions excluding the surface-induced phenomena.

The desired physical conditions naturally occur in the nonmagnetic compounds known as higher manganese silicides (HMS) with the general formula MnSi$_{\gamma}$ ($\gamma = 1.731$--1.750)~\cite{Migas,Miyazaki2008,Weathers}. The HMS share the Nowotny chimney ladder crystal structure with the elongated tetragonal $c$-axis~\cite{Miyazaki2015,Fredrickson,Verchenko}. It is well known that the process of growing bulk single crystals of HMS is inevitably accompanied by the formation of MnSi precipitates in the shape of lamellae that are oriented perpendicular to the [001] crystallographic direction of the HMS matrix~\cite{Aoyama,Suvorova,Miyazaki2018}.

Figure \ref{ris:fig1}(a) shows the (010) cross-section of an oriented Mn$_{11}$Si$_{19}$ single crystal used in the present study. The contrast obtained by polarized-light microscopy in Fig.~\ref{ris:fig1}(a) confirms the presence of MnSi precipitates embedded in the Mn$_{11}$Si$_{19}$ matrix. They are subject to conditions summarized in the Supplemental Material~\cite{Supp}: MnSi in the form of single-crystal lamellae has a mean thickness $l_z \approx 1$~$\mu$m, which is much larger than the spin-spiral period of $\sim 180$~\AA~\cite{Muehlbauer}, and the lateral dimensions $l_x, l_y \gg l_z$, where $l_z, l_x$, and $l_y$ are defined by [001], [100], and [010] directions of the matrix, respectively. The MnSi lamellae are separated by a mean distance $L_z \approx 40$~$\mu$m. At the interface of each lamella, the lattice mismatch between MnSi and Mn$_{11}$Si$_{19}$ produces the tensile stress. Due to the finite Poisson's ratio, in addition to the tensile strain along the interface, it causes a transverse compressive strain of $\sim 1$--3\% in the direction perpendicular to the single-crystalline MnSi lamella, as found in previous transmission electron microscopy studies~\cite{Suvorova}. We note that the actual strain in our sample may differ from the values reported in~\cite{Suvorova} (because of slightly different synthesis procedure). Because the matrix is not magnetic, the magnetization $M(T)$  measured with the commercial SQUID-VSM [Fig. \ref{ris:fig1}(b)] reveals the magnetic ordering temperature $T_{\text C} = 41$~K of the strained MnSi. This is $\sim 30$\% higher than in MnSi without strain~\cite{Supp}. A clear anisotropy of the strained lamellae is indicated by $M$ vs $H$ measurements presented in Fig. \ref{ris:fig1}(c), where a clear metamagnetic transition can be observed for one orientation of the field but not for the other. Hereafter, we use the crystallographic directions of the Mn$_{11}$Si$_{19}$ (tetragonal) to denote the orientation of the applied magnetic field, i.e. $H \parallel [001] \parallel \sigma$ or $H \parallel [100] \perp \sigma$. Figure~\ref{ris:fig1}(d) shows the magnetization and the derivative d$M$/d$H$ data for $T = 35$~K and the field perpendicular to the strain, two transitions at $\sim$~0.15 and 0.35~T are seen.

SANS measurements were carried out at the instrument SANS-1 (FRM-II, Garching)~\cite{SANS-1}. We used a large ($\sim 3$~g) single crystal of  Mn$_{11}$Si$_{19}$ grown as reported elsewhere, which contains in total $\sim 0.1$~g of the strained MnSi lamellae. To explore the stability of the SkL phase, we conducted the SANS experiment in three different configurations: (i) $H \parallel [100]$ with the incident neutrons $n_0 \parallel H$ that brings the $(0KL)$ plane of the matrix in the detector plane, (ii) $H \parallel [100] \perp n_0$ (the $(HK0)$ plane), and  (iii) $H \parallel [001] \perp n_0$. This allowed us to perform the full reciprocal-space imaging of the long-periodic magnetic texture of the strained MnSi and observe the characteristic redistribution of the scattering intensity in different reciprocal-space planes.

\begin{figure}[t]
        \begin{minipage}{0.99\linewidth}
        \center{\includegraphics[width=1\linewidth]{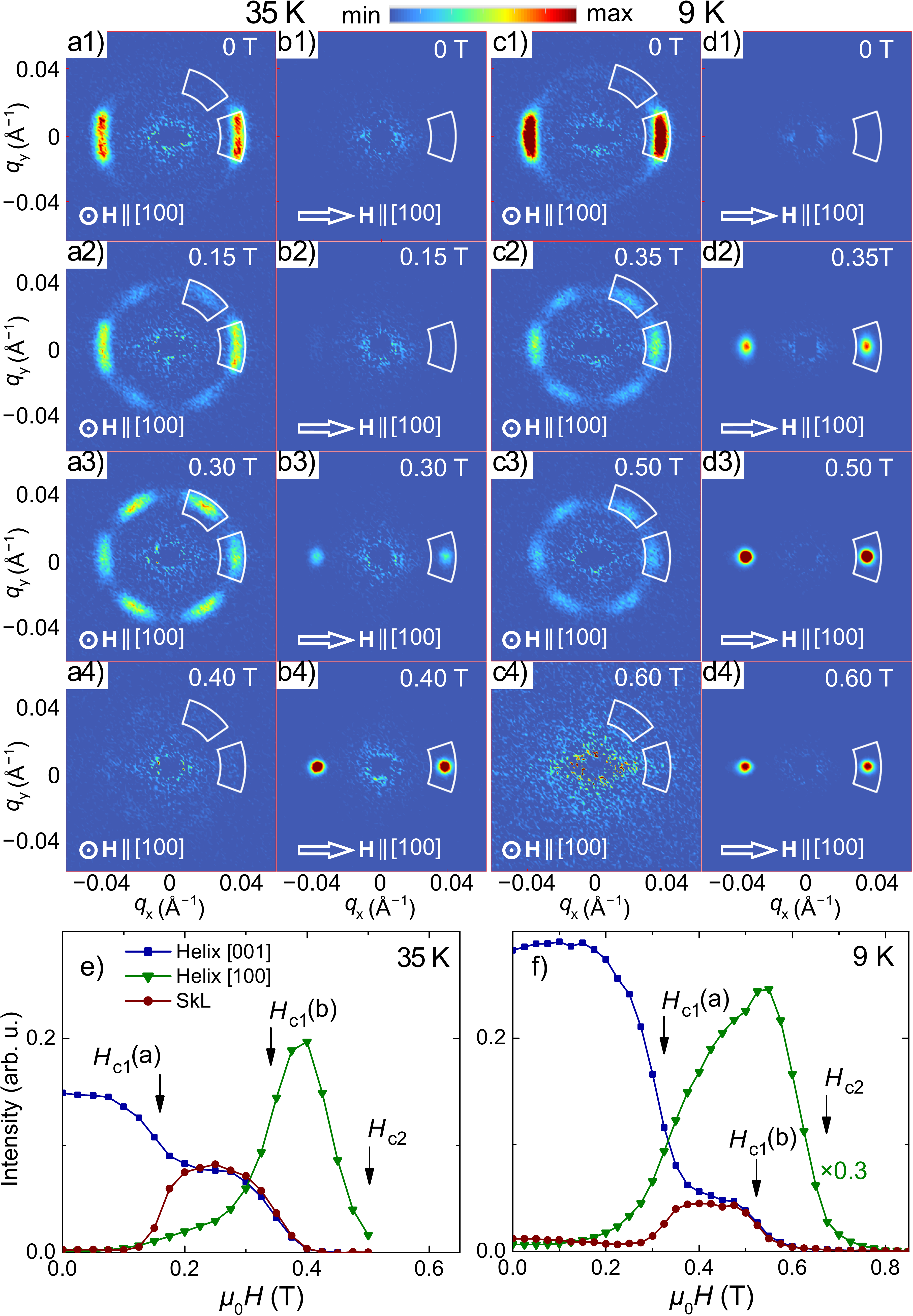}}
        \end{minipage}
        \caption{(color online). SANS patterns recorded at $T = 35$~K (a1)--(b4) and $T = 9$~K (c1)--(d4) in the sample and the applied field geometry as described in the text. The white arrows depict the magnetic field orientation. (e)-(f) SANS intensity integrated within the white sectors in (a1)--(d4) as a function of the field. Solid lines are guides for the eyes. The intensity drawn by green triangles in (f) was multiplied by 0.3 for clarity. The intensities in different configurations are not to scale because of differently broad (larger than the instrumental resolution) magnetic mosaicity of helical, SkL, and conical phases \cite{Supp}.}
        \label{ris:fig2}
\end{figure}

\begin{figure}[t]
        \begin{minipage}{0.99\linewidth}
        \center{\includegraphics[width=1\linewidth]{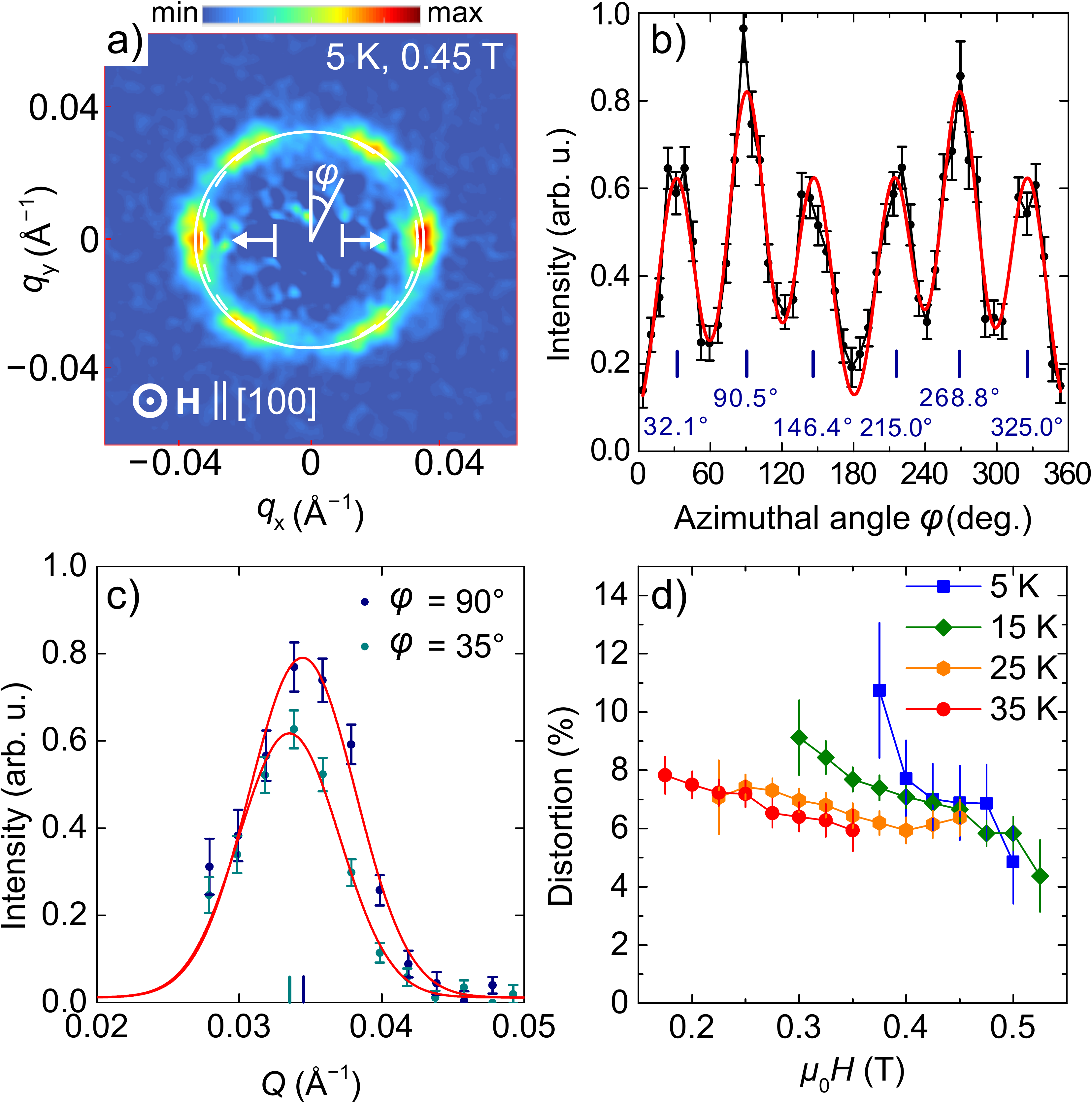}}
        \end{minipage}
        \caption{(color online). The distortion of the SkL in the strained MnSi. (a) SANS pattern collected at $T = 5$~K and $\mu_0H = 0.45$~T, white arrows point to the direction of tension (in the reciprocal space). (b) Intensity of the SkL reflections as a function of $\phi$, solid line is fit by a sum of six Gaussian functions. (c) Radial distribution of the intensity at $\phi = 90^{\circ}$ and $35^{\circ}$, solid lines are Gaussian fit. (d) The observed distortion of the SkL at different temperatures and applied fields, solid lines are guides for the eyes.}
        \label{ris:fig3}
\end{figure}

The results of our SANS measurements in the setups (i) and (ii) for $T = 35$ and 9~K are selectively shown in Fig.~\ref{ris:fig2}. In all the measurements, the sample was zero-field cooled to a fixed temperature, afterwards the magnetic field was applied and increased stepwise. Figures~\ref{ris:fig2}(a1)--\ref{ris:fig2}(b4) demonstrate the SANS maps recorded at 35 K, i.e. 6~K below $T_{\text C}$, in different magnetic fields applied perpendicular to the strain direction. The patterns in Fig.~\ref{ris:fig2}(a1)--(a4) represent the data of the setup (i), whereas Figs.~\ref{ris:fig2}(b1)--\ref{ris:fig2}(b4) is the setup (ii). As can be seen, the helical structure at $H = 0$ forms a single-domain state with the propagation vector $k_s$ oriented along [001] ($\parallel \sigma$). The Bragg peaks are quite sharp in the longitudinal direction (along the momentum transfer $\textbf{Q}$) but somewhat broadened in the azimuthal direction~\cite{Supp}. Upon application of $\mu_0H = 0.15$~T, the six-spot diffraction pattern of the SkL emerges perpendicular to the applied field. However, at 0.15~T the SkL coexists with the spin spiral along [001], as follows from the higher intensity of the $\textbf{Q} \parallel (001)$ reflection. In the field increased to 0.3~T, only the SkL exists in the plane $\perp H$. At the same time, a conical spiral with $k_s \parallel (100) \perp \sigma$ appears [Fig.~\ref{ris:fig2}(b3)]. Finally, at $\mu_0H = 0.4$~T the intensity of SANS is fully redistributes from the SkL observed in the setup (i) to a conical phase [the setup (ii)]. Futhermore, the  exact same features are observed at the low $T = 9$~K in Figs.~\ref{ris:fig2}(c1)--\ref{ris:fig2}(d4), as shown for $\mu_0H = 0$, 0.35, 0.50, and 0.60~T. The precise field evolution of the intensity of different reflections, helical structure with $k_s \parallel (001)$, SkL, and conical with $k_s \parallel (100)$, is depicted in Figs.~\ref{ris:fig2}(e),(f) for the data of 35 and 9~K, respectively. The SkL starts nucleating in the vicinity of the critical field denoted as $H_{\text c1}(a)$. The crossover from a helical state to the SkL is somewhat broad, and the SkL coexists with the helix within a $\sim$ 0.1 T wide region. At the field of $\sim$~0.2~T at 35~K and 0.4~T at 9~K, the intensity of the Bragg peak at $\textbf{Q} = (00k_s)$ becomes equal to other reflections in the SkL plane. At this point only the SkL remains in the plane $\perp H$. The SkL disappears later in the vicinity of $H = H_{\text c1}(b)$, fully transforming to a conical phase oriented along the field. The transition into the field-polarized (FP) state is identified by the vanishing intensity at any finite $Q$. On the contrary, the measurements in the setup (iii), which is for $H \parallel \sigma$, result in trivial observations of the helical-conical and conical-FP transitions~\cite{Supp}. The qualitative analysis of the SANS intensities measured in different configurations and shown in Fig. \ref{ris:fig2} suggests that a finite fraction of the conical phase coexists simultaneously with the SkL. Moreover, the relative contribution of the conical phase might shift with temperature. This possibly originates from a certain distribution of the internal magnetic field in the sample, as was studied in \cite{Reimann}, or/and inherently varying magnitude of the uniaxial strain along the lateral dimension of the lamellae. However, the quantitative description of the possibly varying SkL/conical phase separation remains out of the scope of the present paper.

\begin{figure}[t]
        \begin{minipage}{0.99\linewidth}
        \center{\includegraphics[width=1\linewidth]{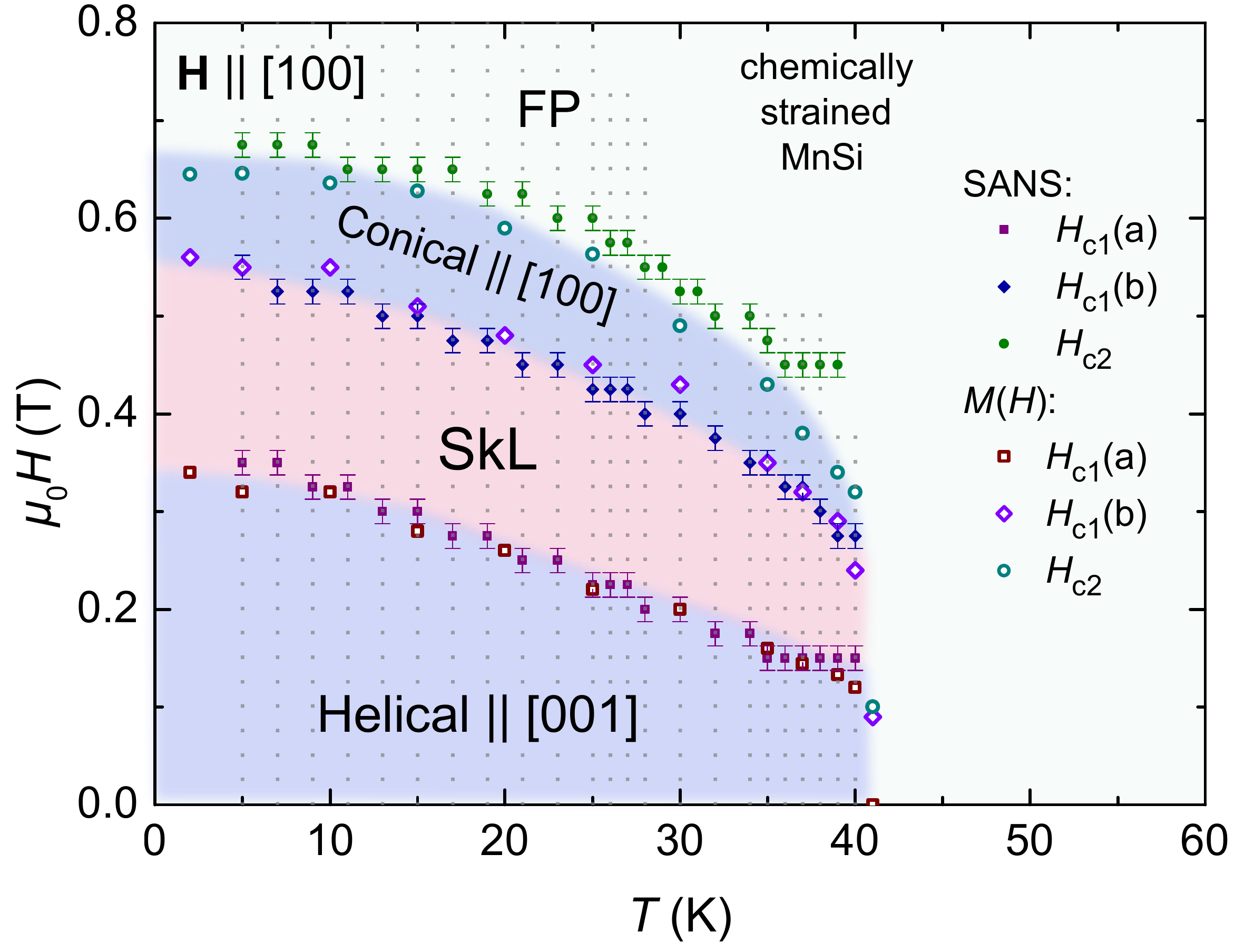}}
        \end{minipage}
        \caption{(color online). The phase diagram of the chemically strained MnSi in the field applied perpendicular to the uniaxial compressive strain. The phase boundaries summarize the results of SANS (filled symbols) and magnetization (opened symbols) measurements in applied magnetic field at constant temperature after ZFC procedure. The red color differentiate the $T$--$H$ region where the stable SkL is observed from the regions where only helical or conical states exists. The mesh of grey dots marks $T$ and $H$ at which SANS patterns have been collected.}
        \label{ris:fig4}
\end{figure}

As seen in Fig.~\ref{ris:fig2}, the Bragg peaks of the SkL do not form a regular hexagon. The emerging distortion of the observed SkL is analysed in Fig.~\ref{ris:fig3}. Figure~\ref{ris:fig3}(a) shows the SANS pattern of the SkL at $T = 5$~K and $\mu_0H = 0.45$~T. The SkL is elongated along the $(00L)$ direction in the reciprocal space. This corresponds to a compression of the SkL in real space, in accordance with the direction of the compressive strain. To quantify the distortion we plotted the intensity as a function of the azimuthal angle $\phi$ [Fig.~\ref{ris:fig3}(b)] and fitted the intensity profile with a combination of six Gaussian functions. In contrast to 60$^{\circ}$-spaced reflections expected for a regular SkL in the chiral magnets free of strain, there exists an azimuthal tilt of $\sim \pm 4^{\circ}$ at $\phi = 90^{\circ} \mp 60^{\circ}$ and $\phi = 180^{\circ} \mp 60^{\circ}$. The distortion can also be resolved in the radial cuts $I = I(Q)$ taken at different $\phi$, as drawn in the Fig.~\ref{ris:fig3}(c). We quantify the distortion $d$ as $d(\%) = 100\% \cdot \langle \Delta \phi\rangle / 60^{\circ}$, where $\langle \Delta \phi\rangle$ is the average tilt of the Bragg peaks toward the $Q \parallel (001)$ line. The resulting values are shown in Fig.~\ref{ris:fig3}(d) as a function of the field at $T = 5, 15, 25$, and 35~K. As follows from the analysis, the SkL distortion tends to decrease in the increasing field from $\sim 8$\% to 6\% at 35 and 25~K, and from $\sim 10$\% to 4\% at lower temperature of 15 and 5~K. It is worth to mention that similar distortions of the SkL were recently observed in strained thin plates of FeGe~\cite{Shibata} and Cu$_2$OSeO$_3$~\cite{Okamura}.

The full $T$--$H$ phase diagram of the chemically strained MnSi in the field applied perpendicular to the compressive strain is presented in Fig.~\ref{ris:fig4}. As was discussed above, combining the SANS data collected in the setups (i) and (ii), one can identify three typical critical fields $H_{\text c1}(a)$, $H_{\text c1}(b)$, and $H_{\text c2}$ that correspond to the transition from the mono-domain helical state with $k_s \parallel (001)$ to the SkL, from the SkL to the conical state with $k_s \parallel (100)$, and to the FP state, respectively. In order to map out the phase diagram, we performed fine field scans at many temperatures and followed the same analysis as presented in Fig.~\ref{ris:fig2}. The critical fields $H_{\text c1}(a)$ and $H_{\text c1}(b)$ that define the boundaries of the SkL were determined as the extrema of the first derivative of the field-dependent SkL intensity $\partial I_{\text{SkL}}/ \partial H$ at constant $T$.   The metamagnetic anomalies seen in the $M(H)$ data [Figs.~\ref{ris:fig1}(c),(d)] are in perfect agreement with the phase transitions observed in SANS. The SkL is present within the field range of $\sim 0.2$~T regardless temperature. The phase diagram constructed from the measurements with $H \parallel [001]$ appears to feature only the topologically trivial conical phase~\cite{Supp}.

To conclude, we have demonstrated that the SkL state can be stabilized in chiral magnets down to the lowest temperature and therefore span the entire $T$-region below the ordering temperature if sufficient strain is applied. Because in the present study the strain has a chemical origin and is not continuously controlled, the strain threshold at which the thermal fluctuations in MnSi become completely irrelevant to the SkL formation cannot be precisely determined. Seemingly, the strain in MnSi inclusions in Mn$_{11}$Si$_{19}$ is significantly larger than that in recent uniaxial-pressure/tension experiments (bulk~\cite{Chacon,Nii,Fobes,Seki17} and thin plates~\cite{Okamura,Shibata}). As follows from our study, the larger strain leads to a significantly enhanced temperature stability of the SkL phase when the magnetic field is applied perpendicular to strain, whereas it gets  fully suppressed for the field along the strain. Supposedly, similar magnitudes of the strain are found in thin films~\cite{Karhu,Sinha,Figueroa}. This might explain the low-temperature anomalies in magnetic~\cite{Wilson12} and magneto-transport measurements~\cite{Gallagher,Yokouchi}. Two physically distinct scenarios have previously been discussed regarding the relation between the strain-induced distortions of the crystal lattice and the change in the energy landscape of magnetic interactions. First is the induced uniaxial single-ion anisotropy~\cite{Butenko,Karhu,Wilson12,Wilson14,Okamura} and the second is the anisotropy of the Dzyaloshinskii-Moriya interaction (DMI)~\cite{Seki17,Shibata}. It is important to note, that the single-ion anisotropy should cause a pronounced difference between the critical fields $H^{\parallel}_{\text c2}$ and $H^{\perp}_{\text c2}$ for the field applied parallel and perpendicular to the anisotropy axis~\cite{Karhu}. We obtained essentially no difference in $H_{\text c2}$ after correcting the data for the demagnetization factor~\cite{Supp}. The observed distortions of the SkL further support the scenario of the anisotropic DMI in accordance with~\cite{Shibata}. The model relation between the anisotropic DMI and the skyrmion stability should be addressed in future theoretical and experimental research.

\textit{Acknowledgements}. A.S.S. acknowledges support from the International Max Planck Research School for Chemistry and Physics of Quantum Materials (IMPRS-CPQM). The work at the TU Dresden was funded by the German Research Foundation (DFG) in the framework of the Collaborative Research Center SFB 1143 (project C03) and the Priority Program SPP 2137 ``Skyrmionics'' (project IN 209/7-1).

\end{document}